\documentclass[transmag]{IEEEtran}
\usepackage{latexsym}
\usepackage{graphicx}
\usepackage{lipsum}
\usepackage{mathtools}
\usepackage{float} 
\usepackage{cuted}
\usepackage{flushend}
\usepackage{booktabs}
\usepackage{multicol}
\usepackage{diagbox}
\usepackage{tabularx}
\usepackage[table,xcdraw]{xcolor}
\usepackage{subfig}
\usepackage{cite}
\usepackage{amsmath,amssymb,amsfonts,mathtools}
\usepackage{algorithm,algcompatible,amssymb,amsmath}

\algnewcommand\algorithmicto{\textbf{to}}
\algnewcommand\RETURN{\State \textbf{return} }
\usepackage{xargs}
\usepackage{textcomp}
\usepackage{xcolor}
\usepackage{multirow}
\usepackage{makecell}
\usepackage{soul}
\usepackage[font=footnotesize]{caption}
\newcommand{\col}{\cellcolor[HTML]{34CDF9}}
\def\BibTeX{{\rm B\kern-.05em{\sc i\kern-.025em b}\kern-.08em T\kern-.1667em\lower.7ex\hbox{E}\kern-.125emX}}
\begin{document}

\title{A Novel Approach for Cancellation of Non-Aligned Inter Spreading Factor Interference in LoRa Systems}

\author{Qiaohan Zhang, Ivo Bizon, Atul Kumar, Ana Belen Martinez, Marwa Chafii, and Gerhard Fettweis
\thanks{This work was supported by the European Union’s Horizon 2020 research and innovation programme through the project iNGENIOUS under grant agreement No.957216, by the German Research Foundation (DFG, Deutsche Forschungsgemeinschaft) as part of Germany’s Excellence Strategy – EXC 2050/1 – Project ID 390696704 – Cluster of Excellence “Centre for Tactile Internet with Human-in-the-Loop” (CeTI) and by the German Federal Ministry of Education and Research (BMBF) (6G-life, project-ID 16KISK001K). We also thank the Center for Information Services and High Performance Computing (ZIH) at TU Dresden.}
\thanks{Qiaohan Zhang, Ivo Bizon, Ana Belen Martinez and Gerhard Fettweis are with the Vodafone Chair Mobile Communications Systems, Technische Universit{\"a}t Dresden (TUD), Dresden, 01187, Germany (emails: \{qiaohan.zhang, ivo.bizon, ana-belen.martinez, gerhard.fettweis\}@tu-dresden.de). Atul Kumar is with the Department of Electronics Engineering, Indian Institute of Technology (BHU), Varanasi, 221005, India (email: atul@iitbhu.ac.in). Marwa Chafii is with the Engineering Division, New York University (NYU) Abu Dhabi, 129188, UAE and with NYU WIRELESS, NYU Tandon School of Engineering, New York, 11201, USA (email:marwa.chafii@nyu.edu).}}

\IEEEtitleabstractindextext{\begin{abstract} Long Range (LoRa) has become a key enabler technology for low power wide area networks. However, due to its ALOHA-based medium access scheme, LoRa has to cope with collisions that limit the capacity and network scalability. Collisions between randomly overlapped signals modulated with different spreading factors (SFs) result in inter-SF interference, which increases the packet loss likelihood when signal-to-interference ratio (SIR) is low. This issue cannot be resolved by channel coding since the probability of error distance is not concentrated around the adjacent symbol. In this paper, we analytically model this interference, and propose an interference cancellation method based on the idea of segmentation of the received signal. This scheme has three steps. First, the SF of the interference signal is identified, then the equivalent data symbol and complex amplitude of the interference are estimated. Finally, the estimated interference signal is subtracted from the received signal before demodulation. Unlike conventional serial interference cancellation (SIC), this scheme can directly estimate and reconstruct the non-aligned inter-SF interference without synchronization. Simulation results show that the  proposed  method  can  significantly  reduce  the  symbol  error rate  (SER)  under  low  SIR compared  with the  conventional  demodulation. Moreover, it also shows high robustness to fractional sample timing offset (STO) and carrier frequency offset (CFO) of interference. The presented  results  clearly  show the  effectiveness  of  the  proposed  method  in  terms  of the SER performance.\end{abstract}


}

\maketitle

\section{Introduction}
\IEEEPARstart{T}{he} internet of things (IoT) enables physical objects to interconnect and exchange data. Long distance and limited power consumption are important requirements for efficient data transmission in IoT. Low power wide area networks (LPWAN) provide a solution to help operators meet the specific coverage and power consumption requirement of IoT.  Long-Range (LoRa) is a popular LPWAN physical layer technology which enables data  communication over a long range while maintaining limited power usage\cite{survey1}. 

LoRa employs chirp spread spectrum (CSS) which is a variant of frequency shift keying (FSK) modulation to encode data \cite{de2021alternative,bomfin2019novel}. LoRa's physical layer standard defines the spreading factor
(SF) as the number of bits that one data symbol represents, which ranges from 7 to 12 bits. Signals modulated with different SFs are assumed to have a very low correlation, i.e., quasi-orthogonality, due to the chirp signal properties \cite{orthogonal}.

When two users with different SFs transmit simultaneously at the same time and frequency, and the source of the desired signal is farther away from the gateway while the source of the undesired signal is closer, the power of the desired signal will be much lower than the undesired signal at the receiving gateway. This problem is known as the near-far problem. In this case, the correlation between the desired and undesired signal is no longer negligible. The undesired signal can interfere with the demodulation, limiting the scalability of LoRa networks\cite{croce2017impact}. In other words, the orthogonality between signals from different SFs is imperfect. The undesired signal can be regarded as non-aligned inter-SF interference, since the signals modulated with different SFs are randomly overlapped. Although the  most  significant  collision occurs when the SF of signal and interference are identical \cite{Impact_interference}, the inter-SF collision cannot be ignored, especially in near-far conditions.  

\subsection{Literature on LoRa performance analysis}

In the current literature, a few works have investigated the issue of collision in same-SF scenario. The investigation in \cite{Solution_ber} derives an approximation for the bit error rate (BER) of the LoRa modulation under additive white Gaussian noise (AWGN) and considering interference from another transmitter with the same SF, whereas the work in \cite{Solution_ser} provides an approximation for the symbol error rate (SER) under a more general non-aligned model. In addition to the mathematical analysis, the authors in \cite{Same_SF_cancel} propose a receiver structure using serial interference cancellation (SIC) technique, which is able to demodulate multiple users simultaneously transmitted over the same frequency channel with the same SF. The work in \cite{2021canc} proposes LoRaSyNc (LoRa receiver with synchronization and cancellation), which can demodulate the strongest signal in case of same-SF collisions, and thus enables signal cancellation and iterative decoding. The authors in \cite{Ben2020} propose an enhanced receiver able to synchronize and decode simultaneously received LoRa signals modulated with the same SF, where commercialized LoRa chips are employed. \textcolor{black}{In \cite{mlora}, mLoRa is proposed for decoding of collided frames modulated with the same SF. It first detects and partly decodes the collision-free samples of the collided frames. The collided samples are then reconstructed and used for SIC. }

Although these publications present appealing results, they do not consider the inter-SF scenario. Furthermore, \cite{Same_SF_cancel}, \cite{2021canc}, \cite{Ben2020} and \cite{mlora} use SIC technique to design LoRa receivers and the performance relies heavily on the synchronization of overlapped signals. The study presented in \cite{croce2017impact} investigates the influence of the imperfect orthogonality between different SFs and shows that  inter-SF collision can degrade LoRa BER performance under low signal-to-interference ratio (SIR). Compared to same-SF interference under which LoRa presents a very high capture probability, inter-SF interference has a different impact on LoRa performance, which can be approximately treated as white noise \cite{afisiadis2020advantage}. 

Furthermore, LoRa PHY coding mechanisms can reduce synchronization error, but cannot mitigate the influence of inter-SF collisions \cite{Impact_interference}.  LoRa employs a discrete Fourier transform (DFT) based mechanism for time and frequency synchronization \cite{bernier2020low}, \cite{xhonneux2019low}, which can also be affected by inter-SF collision. In \cite{Croce2020} it is shown that high SFs are severely affected by inter-SF collision, and using power control and packet fragmentation produces very little effect on this problem. The authors in \cite{Interference_analysis} further extend the study of inter-SF interference to investigate the influence of the chirp rate, which is dependent on both SF and bandwidth. Their results show agreement with the values of SIR threshold to those reported in \cite{croce2017impact}. 

Nevertheless, the issue of inter-SF collision has been seldom mathematically analyzed in the literature. Compared to same-SF interference, inter-SF collision is more difficult to model and estimate. First, the  SF  of  the interference is unknown at the LoRa gateway and thus needs to be identified. Furthermore, the detection and synchronization of the interference signal can be prevented if the received signal does not contain the preamble, or the preamble structure is unknown.  

\subsection{Contributions}	
In this work, we model the inter-SF interference in LoRa, and propose a method that enables the receiver to estimate and reconstruct the non-aligned inter-SF interference from another unknown user. Therefore, allowing interference cancellation at the gateway. Our main contributions are:
\begin{itemize}
    \item An algorithm to estimate the equivalent non-aligned inter-SF interference signal, including the SF identification, equivalent data symbol, and complex amplitude estimation, which therefore enables the cancellation of interference.
    \item We evaluate the SER, BER of perfectly synchronized signals in the presence of interference under AWGN channels, and show that this method can significantly improve the demodulation performance.
\end{itemize}

\subsection{Organization}
This paper is organized as follows. Section II provides an overview of the system model of LoRa, where the mechanism of CSS modulation is also introduced. In Section III, the modeling of inter-SF interference is presented. The proposed interference cancellation scheme is illustrated step by step. Section IV shows the simulation results of the proposed algorithm. In the end, conclusions are drawn in Section V. The notations and symbols used in this work are defined in Table \ref{tab:CNN}.
\begin{table}[t!]
\centering
\captionsetup{justification=centering}
\caption{Notations and symbols definition.
}
\begin{tabular}{@{}l l@{}}
\toprule
Symbol          & Definition \\ \midrule
$x_k[n]$ &Desired LoRa signal with data symbol $k$     \\
$\mathrm{SF}_u, \mathrm{SF}_i$ &Spreading factor of user and interferer\\
$A_u, A_i$         & Signal amplitude of user and interferer    \\
$N_u, N_i$          &\begin{tabular}[c]{@{}l@{}}Number of samples in one LoRa signal of user and interferer\end{tabular}    \\
$c[n]$           & Raw up-chirp of $N_u$    \\
$y_k[n]$ &Received LoRa signal with desired data symbol $k$     \\
$h_u, h_i$     &Complex valued channel gain of desired signal and interference     \\
$l$ &Data symbol of interference\\
$\tau$ &Sample timing offset of interference\\
$\chi$          &Carrier frequency offset of interference          \\
$i_l\lbrack n+\tau\rbrack$       &Interference signal with data symbol $l$ and sampling offset $\tau$\\
$w[n]$       &Additive white Gaussian noise\\
$l_q$       &$q$-th interference symbol collided with the desired symbol\\
$r[n]$        &Despread received signal           \\
$R[f]$         & $N_u$-point DFT of despread received signal \\
$\widehat {\mathrm{SF}}_i,\ \widehat l,\ \widehat A_i $  &Estimated $ {\mathrm{SF}}_i,\  l,\  h_i A_i $           \\
$\phi$        &Phase of $h_i A_i$\\
$\widehat \phi$        &Phase of $\widehat A_i$\\
$\widehat i_l[n]$        &Reconstructed interference signal  \\
$\tilde{y}_k[n]$        &Received signal after interference cancellation    \\
$N$          &Possible $N_i$        \\
$R_N[f]$          &$N$-point DFT of despread received signal \\
$C[N]$ &Peak-to-average ratio of the absolute value of $R_N[f]$\\
$c^*_N[n]$        &Raw down-chirp of $N$\\
$r_{N}[n]$ &Despread signal using a down-chirp modulated with $\mathrm{SF}=\log_2N$\\
$P_{\mathrm{DFT}}$ &Number of DFT blocks\\
$N_s$          &Number of samples for interference SF identification        \\
$d$ &Number of segments of a interference signal\\
$\theta$        &Symbol shift induced by $\tau$ and $\chi$\\
$i_l^m\left[n\right]$ &$m$-th segment of interference signal $i_l\lbrack n+\theta\rbrack$\\
$c_{N_i}^*[n]$ &Raw down-chirp of $N_i$\\
$s_l^m\left[n\right]$ &Despread signal of $i_l^m\left[n\right]$\\
$\Phi$ &Phase shift induced by $\phi$ and $\tau$\\
$Q_l^m[f]$ &DFT of $s_l^m\left[n\right]$\\
$\widehat l_m$ &Estimated equivalent data symbol of $i_l^m\left[n\right]$\\
$\tilde{i}_l\left[n\right]$ &Reconstructed interference signal without amplitude estimation\\
$i_{l,\tau,\chi}[n]$ &Received interference signal\\
$\widehat A^m_i$ &Estimation of $A_ie^{j\Phi}$ for the $m$-th segment\\
$\chi_{frac},\tau_{frac}$ &Fractional $\chi$ and $\tau$\\
$ |\widehat A_i^m|_d$ &Magnitude of $\widehat A^m_i$ using $d$ segments\\
$d_{best}$ &$d$ value that shows best interference estimation performance\\
$M_d$ &Mean square error of $ |\widehat A_i^m|_d$\\
$V_d$ &Variance of $ |\widehat A_i^m|_d$\\
$|\cdot|$ &Operation of absolute value\\
max($\cdot$) &Operation of maximum value\\
min($\cdot$) &Operation of minimum value\\
$\lfloor\cdot\rfloor$ &Operation of floor\\
\bottomrule
\end{tabular}\label{tab:CNN}
\end{table}
	
	\begin{figure*}[ht!]
		\centering
		\includegraphics[]{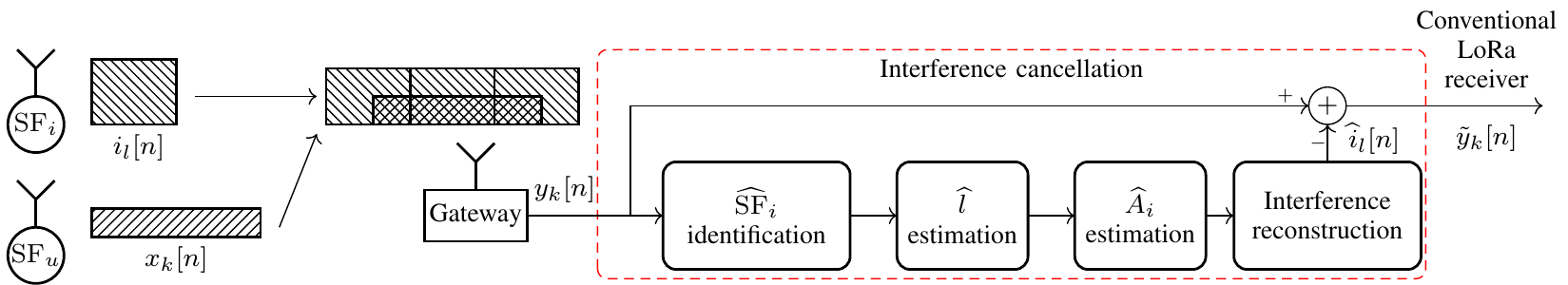}
		\caption{Block diagram of the proposed interference cancellation scheme. The desired signal is added to a signal with a different SF. At the receiver side, the interference cancellation is performed on the signal. The conventional demodulation is employed after interference cancellation.}
		\label{LoRATXRX}
	\end{figure*}
\section{System Model}\label{sec:System_model}
 The CSS modulation employed in LoRa is implemented by using linear frequency modulated chirp pulses to encode information, which enables the energy to spread in frequency \cite{Lora_modu}. We consider a  LoRa receiver
sampling signals at the Nyquist frequency. Then, the equivalent equation of
a discrete LoRa signal in the physical layer (PHY) is given by \cite{Ivo}:
 	\begin{equation}\label{eq:Lora_symbol}
	x_k[n] = A_u \exp \left( j\frac{2\pi }{N_u}kn \right) c\lbrack n\rbrack,
	\end{equation}
 where $N_u = 2^{\mathrm{SF}_{u}}$ is the total number of samples within one LoRa signal, $\mathrm{SF}_u$ ranges from 7 to 12 and represents number of bits contained in one data symbol, $A_u=\sqrt{{E_s}/{N_u}}$ represents the signal amplitude, where $E_s$ is the symbol energy, and $n$ is the sample index.
The complex exponential contains the data symbol $k$ and its energy is spread by the discrete-time raw up-chirp  $c\lbrack n\rbrack = \exp\left( j\pi n^2/N_u \right)$. The data symbols are integer values from the set $ \mathbb{K} = \left\lbrace 0, \ldots, \, 2^{\mathrm{SF}_{u}} - 1\right\rbrace $. The bit-words containing SF bits are mapped onto one symbol $k$, which feeds the CSS modulator.  

We specify the case in which only one interferer using a different SF exists. {In addition, we also consider that the user and the interferer operate using the same bandwidth.} Assuming  a  receiver  perfectly synchronized in time and frequency, \textcolor{black}{a data symbol interfered with data symbols from another user modulated with a different SF at the gateway can be described by \eqref{eq:signal_inter}}, 
where $i_l\lbrack n+\tau \rbrack$ represent interference signals. $A_i$ is the amplitude and \textcolor{black}{$l \in \mathbb{L} = \left\lbrace 0, \ldots, \, 2^{\mathrm{SF}_{i}} - 1\right\rbrace $} are the interference data symbols. $N_i = 2^{\mathrm{SF}_i}$ represents the total number of samples within one interference data symbol, where $\mathrm{SF}_i$ is the SF of the interferer, $n+\tau$ is the sample index of the interference and $\tau$ is the sample timing offset (STO) between the desired signal and interference, $e^{j2\pi\chi n/{N_i}}$ is the phase shift induced by $\chi$ which is the carrier frequency offset (CFO) of the interference signal, $h_u$ and $h_i$ are the complex-valued channel gains experienced by the desired user and interferer, respectively. {$w\lbrack n\rbrack\sim\mathcal C\mathcal N(0,\,\sigma^2)$} is AWGN. To be more precise, since $N_u\neq N_i$, one desired data symbol can overlap multiple interference data symbols in serial, or vice versa. \textcolor{black}{The complete expression of $i_l\lbrack n+\tau\rbrack$ is shown in \eqref{eq:combi}, where $l_q$, $q\in \mathbb{Z} $, is the $q$-th interference symbol collided with the desired symbol.}
\begin{figure*}[t]
	\begin{equation}\label{eq:signal_inter}
	\begin{split}
y_k\lbrack n\rbrack&=h_u x_k\lbrack n\rbrack+h_i i_l\lbrack n+\tau\rbrack e^{j2\pi\chi n/{N_i}}+w\left[n\right]\\
&=h_u A_u\exp\left(j\frac{\pi} {N_u}\left(n^2+2kn\right)\right)+h_i A_i\exp\left(j\frac{\pi} {N_i}\left(n^2+2n\left(l+\tau+\chi\right)\right)\right)\exp\left(j\frac{\pi} {N_i}\left(\tau^2+2l\tau\right)\right)+w\left[n\right],
	\end{split}
	\end{equation}
\begin{subequations}
\raggedleft
\begin{flalign}
\mathrm{if}\;N_i<N_u\label{eq:inter_1}\mathrm{:}&\nonumber\\&i_l\lbrack n+\tau\rbrack=\left\{\begin{array}{l}\exp \left( j\frac{2\pi l_0\left(n+\tau\right) }{N_i} \right)\exp\left( j \frac{\pi \left(n+\tau\right) ^2}{N_i} \right),\;\mathrm{for}\;n=0,\dots,\;N_i-1-\tau\;,\\\exp \left( j\frac{2\pi l_1\left(n-N_i+\tau\right)}{N_i} \right)\exp\left( j \frac{\pi \left(n-N_i+\tau\right) ^2}{N_i} \right),\;\mathrm{for}\;n=N_i-\tau,\dots,\;2N_i-1-\tau,\\\dots\\\exp \left( j\frac{2\pi l_{{N_u}/{N_i}}\left(n-N_u+\tau\right)}{N_i} \right)\exp\left( j \frac{\pi \left(n-N_u+\tau\right) ^2}{N_i} \right) ,\;\mathrm{for}\;n=N_u-\tau,\dots,\;N_u-1,\end{array}\right.\;\\
    \mathrm{if}\;N_i>N_u\;&\mathrm{and}\;\tau\leq N_i-N_u\mathrm{:}\ \ \ \ \ \ \  &\nonumber\\&i_l\lbrack n+\tau\rbrack=\exp \left( j\frac{2\pi l_0\left(n+\tau\right) }{N_i} \right)\exp\left( j \frac{\pi \left(n+\tau\right) ^2}{N_i} \right),\;\mathrm{for}\;n=0,\dots,\;N_u-1,\;\label{eq:inter_22}\\
\mathrm{if}\;N_i>N_u\;&\mathrm{and}\;\tau>N_i-N_u\mathrm{:}\ \ \ \ \ \ \  &\nonumber\\&i_l\lbrack n+\tau\rbrack=\left\{\begin{array}{l}\exp \left( j\frac{2\pi l_0\left(n+\tau\right) }{N_i} \right)\exp\left( j \frac{\pi \left(n+\tau\right) ^2}{N_i} \right),\;\mathrm{for}\;n=0,\dots,\;N_i-1-\tau\;,\\\exp \left( j\frac{2\pi l_{1}\left(n-N_i+\tau\right)}{N_i} \right)\exp\left( j \frac{\pi \left(n-N_i+\tau\right) ^2}{N_i} \right),\;\mathrm{for}\;n=N_i-\tau,\dots,\;N_u-1.\end{array}\right.\;\label{eq:inter_3}
\end{flalign}
\label{eq:combi}
\end{subequations}
\hrule
\end{figure*}

In the conventional LoRa demodulation scheme, the received signal $y_{k}[n]$ is first multiplied by the down chirp $c^{*}[n]$ to obtain
\begin{equation}
    r[n]= y_k[n]\mathrm{exp}\left(-j\pi n^2/{N_u}\right).
\end{equation}
Then, the non-coherent detection in LoRa is performed by selecting the frequency index with the maximum magnitude, which enables the demodulation without compensation of channel phase rotation. This operation can be described as
	\begin{equation}\label{eq:abs_FFT}
	\hat{k} = \arg\max_{f \in \mathbb{K}} \big| R[f]  \big| ,
	\end{equation}
	where $R[f] = \mathcal{F}\left\lbrace r[n]\right\rbrace$, and $\mathcal{F}\left\lbrace \cdot \right\rbrace$ represents the DFT operator, which can be easily performed via the fast Fourier transform (FFT) algorithm. 
\vspace{-0.3cm}
\section{Interference Modeling}\label{Sec:Inter_eli}
 This section presents a method to estimate and cancel the interference signal modulated with a different SF without any prior knowledge of the interference. \textcolor{black}{We consider an interference signal from a single interferer modulated with another unknown SF, which overlaps the desired signal randomly.} Let $\widehat i_l[n]$ be the reconstructed interference signal, where $\widehat N_i=2^{\widehat{\mathrm{SF}}_i}$ is the estimated $N_i$ of the interference, $\widehat l$ and $\widehat A_i$ are the estimated equivalent interference data symbol and amplitude, respectively. {Note that $\widehat A_i=|\widehat A_i| e^{j\widehat\phi}$ is the estimation of $A_i |h_i|e^{j\phi}=h_i A_i$ which is a phasor. Therefore, both the magnitude $A_i|h_i|$ and the phase $\phi$ of it should be estimated. } The signal after interference cancellation is given by
	\begin{equation}
	\begin{split}
\tilde{y}_k\lbrack n\rbrack&=y_k\lbrack n\rbrack-{\widehat i_l}\left[n\right]\\&=y_k\lbrack n\rbrack-|\widehat A_i|\exp\left(j\frac\pi{\widehat N_i}\left( n^2+2\widehat l n\right)+ j\widehat \phi\right).
	\end{split}\label{eq:Inter_canc}
	\end{equation}
In this section, the algorithm for estimating $\mathrm{SF}_i$ is firstly presented. Then, the estimators of $l$ and $h_iA_i$ are respectively introduced. The interference cancellation is presented in the end. The entire block diagram of the interference cancellation scheme is shown in Figure \ref{LoRATXRX}. 
\subsection{Interference signal SF identification}\label{sec:iden}
The LoRa signals modulated with different SFs have the quasi-orthogonality property. Suppose a LoRa signal modulated with $\mathrm{SF}_1$. If this signal is despread and demodulated using $\mathrm{SF}_2$,
where $\mathrm{SF}_1 \neq \mathrm{SF}_2$, there will not be a prominent peak of the DFT output. In other words, the magnitude of an $N$-point DFT of a despread signal modulated with $\mathrm{SF}_i$ will show a single prominent peak only when $N=2^{\mathrm{SF}_i}$.  Although the non-alignment of the interference signals results in two attenuated peaks within the frequency spectrum, as shown in Figure \ref{fig:peak_2}, the signal energy is not spread. Therefore, this characteristic can be employed for ${\mathrm{SF}}_i$ identification. However, under a noisy environment, the DFT with unmatched SFs can show multiple peaks and the highest one can be higher than that with the matched SF. Hence, instead of using the peak of DFT output, we employ the peak-to-average ratio (PAR) of the DFT magnitude as a metric to estimate ${\mathrm{SF}}_i$. The PAR is given by
\begin{equation}\label{eq:par}
    C\lbrack N\rbrack= \frac{\big| R_N\lbrack f\rbrack\big|_{max} }{\frac{1}{N}\sum_{f=0}^{N-1}\big| R_N\lbrack f\rbrack\big|},
\end{equation}
where $R_N\lbrack f\rbrack$ is the $N$-point DFT of $r_{N}[n]$ which is the despread signal using a down-chirp $c^*_N[n]=\mathrm{exp}\left(-j\pi n^2/{N}\right)$ and $N \in \mathbb{T}=\left\{2^7,\ \dots,\  2^{12}  \right\}\setminus\left\{2^{\mathrm{SF}_u}\right\}$ representing the set of all possible $N_i$.
Then, $\widehat {\mathrm{SF}}_i$ can be obtained via $C\lbrack N\rbrack_{max}$, which is the maximum value of $C\lbrack N\rbrack$. It is clear that the required number of samples from the received signal for an $N$-point DFT block must equal to $N$. When the number of the provided samples $N_s$ for interference SF identification differs from $N$, two possibilities arise. In other words, when an $N$-point DFT block is present, there are $N_s-N$ extra samples if $N<N_s$, whereas in the other case the DFT cannot be performed, since $N>N_s$ and $N-N_s$ samples are missing. In order to solve these problems, the proposed algorithm is subdivided into two cases.
\paragraph{$N\leqslant N_s$}
The signal length is equal or larger than the DFT block, and the DFT can be performed multiple times in series. Thus, let $P_{\mathrm{DFT}}=\lfloor N_s/N\rfloor$ be the total amount of $N$-point DFT blocks within $N_s$ samples. The $N$-point DFT is performed for $P_{\mathrm{DFT}}$ times in sequence on the signal. Each DFT block presents a PAR and only the largest of all PARs is chosen as ${C\lbrack N\rbrack}$.
\paragraph{$N>N_s$}
The DFT block size exceeds the number of the provided samples, and the DFT cannot be performed. The zero-padding method can be applied to solve this, where we pad $N-N_s$ zeros at the end of the signal. Then, a single-block $N$-point DFT can be performed. The PAR can be calculated using \eqref{eq:par}.

$C\lbrack N\rbrack$ is then calculated for the range of possible $N$. $\widehat {\mathrm{SF}}_i$ is obtained from $\widehat N_i$, which can be expressed as
\begin{equation}
    \widehat N_i=\arg\max_{N \in \mathbb{T}} C\lbrack N\rbrack.
    \vspace{-.2cm}
\end{equation}
Algorithm \ref{al:id} illustrates the proposed interference signal SF estimation. It is noted that using an $N_s$ as large as possible reduces the estimation error by evaluating the results of several blocks of DFT.
\begin{algorithm}[t!]
 \caption{Algorithm for $\mathrm{SF}_i$ identification}\label{al:id}
 \begin{algorithmic}[1]
 \renewcommand{\algorithmicrequire}{\textbf{Input:}}
 \renewcommand{\algorithmicensure}{\textbf{Output:}}
 \REQUIRE Received signal $y_k[n]$ with $N_s$ samples 
 \ENSURE  $\widehat {\mathrm{SF}}_i$
\STATE $ N_i=\left[2^7;\ 2^8;\ 2^9;\ 2^{10};\ 2^{11};\ 2^{12}  \right]\setminus2^{\mathrm{SF}_u}$ 
  \FOR {$z=1:5$ }
  \STATE $N=N_i[z]$
  \IF {($N \leq N_s$)}
  \STATE $P_{\mathrm{DFT}}\gets \mathrm{floor}(N_s/N)$
  \FOR {$a=1:P_{\mathrm{DFT}}$}
  \STATE $r_{N}[n]\gets y_k[(a-1)*N+1:a*N]$ despreading
  \STATE $|R_N\lbrack f\rbrack\big|\gets\left|\mathrm{FFT}(r_{N}[n])\right|$
  \STATE $\mathrm{PAR}(a)\gets\max (|R_N\lbrack f\rbrack\big|)/{\mathrm{mean}\left(|R_N\lbrack f\rbrack\big|\right)}$
  \ENDFOR
  \STATE $C\lbrack N\rbrack\gets\max(\mathrm{PAR})$
  \ELSE
  \STATE $y_k[n]\gets (N-N_s)$ zero-padding
  \STATE $r_N[n]\gets y_k[n]$ despreading
  \STATE $|R_N\lbrack f\rbrack\big|\gets\left|\mathrm{FFT}(r_N[n])\right|$
  \STATE $C\lbrack N\rbrack\gets\max (|R_N\lbrack f\rbrack\big|)/{\mathrm{mean}\left(|R_N\lbrack f\rbrack\big|\right)}$
  \ENDIF
  \ENDFOR
  \STATE $\widehat {\mathrm{SF}}_i \gets \widehat N_i \gets z \gets \max(C\lbrack N\rbrack)$
 \end{algorithmic} 
 \end{algorithm}

\subsection{Estimation of the data symbol from the interference signal}\label{sec:symbol_value}
\begin{figure}[tbp]
\captionsetup{width=.5\linewidth}
\subfloat[The boundary between the LoRa signals around the middle of the DFT block.]{
        \includegraphics[]{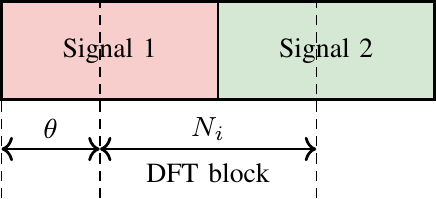}
    \label{fig:simu1}}
\captionsetup{width=.47\linewidth}\subfloat[DFT magnitude assuming non-aligned demodulation of two consecutive data symbols.]{
	\includegraphics[]{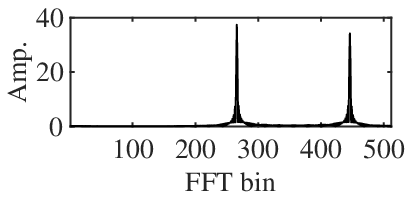}\label{fig:peak_2}}
\captionsetup{width=1\linewidth}
		\caption{Non-aligned demodulation
		of interference signal, where $N_i=512$.}
		\label{non_align}
		\vspace{-0.5cm}
	\end{figure}

\begin{figure}[t]
		\centering
		\includegraphics[]{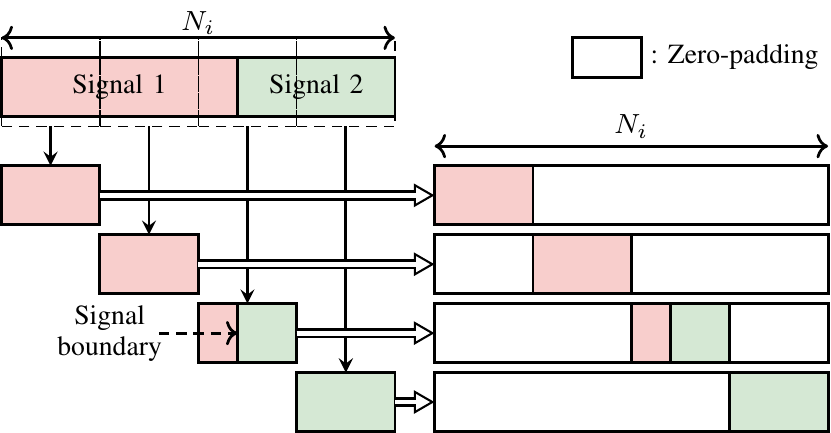}
		\caption{Segmentation of interference signal with $d$=4. The signal with the length $N_i$ is divided into $d$ parts and padded with zeros.}
		\label{Inter_demodu_optimized}
		\vspace{-0.5cm}
	\end{figure}

\begin{figure}[t!]
		\centering
		\includegraphics[width=0.5\textwidth]{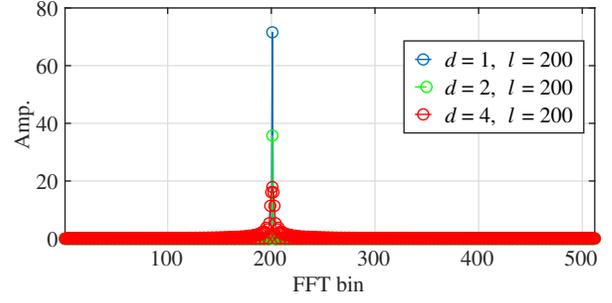}
		\caption{DFT spectrum of a synchronized LoRa signal modulated with SF=9 with segmentation. }
		\label{non_align_symbol_esti}
		\vspace{-0.5cm}
	\end{figure}

{As the interference signal is not synchronized,  direct estimation of the data symbol is not possible due to symbol shift.} However, it is sufficient to estimate the equivalent interference signal under unsynchronized condition for performing interference cancellation. In this case, an $N_i$-point DFT block likely contains samples from two consecutive interference symbols. 
An intractable problem appears when the boundary between two LoRa signals is at the middle of the DFT block, as shown in Figure \ref{non_align}, where the DFT result shows two comparable peaks, but only the index of the largest will be selected. This results in the loss of the data symbol represented by the smaller peak.
Segmentation of the received signal can mitigate the data loss caused by non-alignment, since the information is spread into each sample of the LoRa signal due to the energy spreading. Any segment of the LoRa signal contains the information to identify the data symbol. In other words, non-aligned interference signals can be equivalently estimated via its segments at the expense of reduced processing gain, but with robustness against non-alignment. Each interval of the received signal with the length $N_i$ can be divided into $d\in\mathbb{D}=\left\{1, 2, 4, 8,\dots, N_i\right\}$ segments. In order to perform DFT on the segments extracted from the signal with the length $N_i/d$, zero-padding is employed to fill the vacant sample positions to obtain again a signal with $N_i$ samples, as shown in Figure \ref{Inter_demodu_optimized}. It is worth noting that after segmentation and zero-padding only one segment of the signal involves the information of different data symbols, while the rest of the segments only contain the information of a single data symbol that can be accurately estimated. {The latter scenario is chosen for the following analysis.} We denote $\theta=\tau+\chi$. The $m$-th segment of an interference signal padded with zeros can be expressed as
\begin{equation}\label{eq:segement}
\\i_l^m\left[n\right]\hspace{-0.08cm}=\hspace{-0.08cm}\left\{\begin{array}{l}\hspace{-0.08cm}0,\;\;\;\;\mathrm{for}\;n=0,\dots,\  N_im/d-1\;,\\\hspace{-0.08cm}A_i|h_i|\mathrm{exp}\left(j\frac\pi{{{}_N}_i}\hspace{-0.08cm}\left(n^2+2n\left(l+\theta\right)+\tau^2+2l\tau\right)+j\phi\right),\;\\\;\;\;\;\;\;\hspace{-0.4cm}\mathrm{for}\;n=N_im/d,\dots,\  N_i\left(m+1\right)/d-1,\\0,\;\;\;\;\mathrm{for}\;n=N_i\left(m+1\right)/d,\dots ,\ N_i-1,\end{array}\right.\\
\end{equation}
where $m\in\left[0, d-1 \right]$. Multiplying \eqref{eq:segement} by the down chirp $c_{N_i}^*[n]=e^{-j\pi\frac{n^2}{N_i}}$ yields

\begin{equation}\label{eq:segement_despread}
s_{l}^m[n] =\left\{\begin{array}{l}0,\;\;\;\mathrm{for}\;n=0,\dots,\; N_im/d-1\;,\\A_i|h_i|\exp\left(j\frac{2n\pi}{Ni}\left(l+\theta\right)+j\Phi\right),\;\\\mathrm{for}\;n=N_im/d,\dots,\;N_i\left(m+1\right)/d-1,\\0,\;\;\;\mathrm{for}\;n=N_i\left(m+1\right)/d,\dots,\ N_i-1.\end{array}\right.\\
\end{equation}
This expression indicates that each despread segment contains the equivalent data symbol value $\lfloor l+\theta \rfloor \mod N_i$ and a phase offset $e^{j\Phi}$, where $\Phi=\pi (\tau^2+2l\tau)/N_i+\phi$. An example is shown in Figure \ref{non_align_symbol_esti}. As one can see, the peak of DFT magnitude is inversely proportional to $d$. But the  frequency bin of DFT using segmentation remains the same. Although the segmented demodulation induces symbol shift due to $\theta$, the correct data symbol of the interference is not of interest. After zero padding, each segment contains the information for extracting the equivalent data symbol of interference, which can be expressed as 
\begin{equation}\label{eq:lm}
        \widehat l_m=\arg\max_{f \in \mathbb{L}} \big| Q_l^m[f]  \big|,\;\;\mathrm{for}\;m=0,\dots,\,d-1,
\end{equation}
where $Q_l^m[f] = \mathcal{F}\left\lbrace s_l^m\left[n\right]\right\rbrace$ and $\widehat l_m$ is the estimation of ${\lfloor l+\theta\rfloor \mod\ N_i}$. Then, the interference signal can be reconstructed by extracting the desired data from $d$ estimated segment signals, which can be expressed as
\begin{equation}\label{eq:recover}
\\\tilde{i}_l\left[n\right]=\left\{\begin{array}{l}\mathrm{exp}\left(j\frac\pi{{{}_N}_i}\left(n^2+2n{\widehat l_0}\right)\right),\;\\\;\;\;\;\;\;\mathrm{for}\;n=0,\dots,\   N_i/d-1\;,\\\mathrm{exp}\left(j\frac\pi{{{}_N}_i}\left(n^2+2n{\widehat l_1}\right)\right),\;\\\;\;\;\;\;\;\mathrm{for}\;n=N_i/d,\dots,\   2N_i/d-1,\\\dots\\\mathrm{exp}\left(j\frac\pi{{{}_N}_i}\left(n^2+2n{\widehat l_{d-1}}\right)\right),\;\\\;\;\;\;\;\;\mathrm{for}\;n=N_i\left(d-1\right)/d,\dots,\ N_i-1.\end{array}\right.\\
\end{equation}
Comparing \eqref{eq:signal_inter} with \eqref{eq:recover} yields to
\begin{equation}\label{eq:difference}
\tilde{i}_l\left[n\right]\approx \frac{ i_{l,\tau,\chi}\left[n\right]}{A_i|h_i|e^{j\Phi}},
\end{equation}
where $ i_{l,\tau,\chi}\left[n\right]= h_i i_{l}\lbrack n+\tau\rbrack e^{j2\pi\chi n/{N_i}}$ is the received interference signal. It can be recognized that it is necessary to still estimate $A_i|h_i|$ and $\Phi$ to reconstruct the interference signal.
\subsection{Estimation of the complex amplitude of the interference signal}
The goal is to estimate the complex amplitude of the interference signal without knowledge of SIR. For a perfectly synchronized receiver under noise-free condition, assuming a desired LoRa signal modulated with $\mathrm{SF}_u$ carries a data symbol $k$. Its DFT magnitude yields a peak when $f=k$, which can be described as
	\begin{equation}
	\begin{split}
	\max_{f \in \mathbb{K}} \big| R\left[f\right]\big| &=\big|R[k]\big|\\&=\left| 
	\sum_{n=0}^{N_u-1}\exp(-j2\pi\frac{kn}{N_u})\cdot h_uA_u\exp\left(j2\pi\frac{kn}{N_u}\right)\right|\\&=N_uA_u|h_u|.
	\end{split}
	\end{equation}
Thus, 
\begin{equation}
    A_u|h_u|=	\max_{f \in \mathbb{K}}\big| R\left[f\right]\big|/N_u.
\end{equation}
Similarly, from \eqref{eq:segement_despread} and \eqref{eq:lm} we can obtain
\begin{equation}
 \max_{f \in \mathbb{L}} \big| Q_l^m\left[f\right]\big| =\left |Q_l^m[\widehat l_m]\right|= \left | \frac{N_{i}}{d}A_i|h_i|e^{j\Phi} \right|,\label{eq:Peak_seg}
\end{equation}
where $N_i/d$ is the segment length of the interference signal. Note that the exponential term is the phase of the interfering signal. Therefore, the complex value of \eqref{eq:Peak_seg} is chosen for estimation of the complex amplitude of the interference signal. Thus,
\begin{equation}
\begin{split}
        \widehat A^m_i =  \frac{d}{N_{i}}Q_l^m[\widehat l_m], 
\end{split}\label{eq:amplitude}
\end{equation}
which is the estimation of $A_i|h_i|e^{j\Phi}$ for the $m$-th segment. Similarly to \eqref{eq:recover}, $\widehat A_i$ can be expressed as a piecewise function of $n$
\begin{equation}
\\\widehat A_i\left[n\right]=\left\{\begin{array}{l}\widehat A^0_i,\;\mathrm{for}\;n=0,\dots,\   N_i/d-1\;,\\\widehat A^1_i,\;\mathrm{for}\;n=N_i/d,\dots,\   2N_i/d-1,\\\dots\\\widehat A^{d-1}_i,\;\mathrm{for}\;n=N_i\left(d-1\right)/d,\dots,\ N_i-1.\end{array}\right.\\
\end{equation}
Then, we have the interference signal reconstructed as
\begin{equation}
     \widehat i_l\left[n\right]=\widehat A_{i}\left[n\right] \tilde{i}_l\left[n\right].
\end{equation}
{We note that this estimation is valid only for the segment that contains the information of one LoRa data symbol, i.e., same-symbol segments. The inaccurate interference estimation will happen  when the segment contain the boundary of two consecutive signals, i.e., inter-symbol segment. However, this issue can be mitigated using smaller segments.}

\subsection{Interference Cancellation}
\begin{figure}[t!]
    \centering
    \subfloat[The interference signal includes only integer $\chi$ and $\tau$.]{
        \includegraphics[]{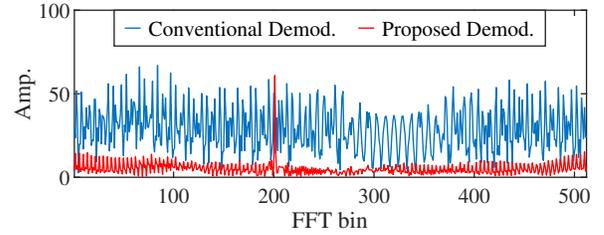}\label{fig:FFT1}
    }
    
    \subfloat[The interference signal includes $\chi$ and $\tau$, where $\chi_{frac}=0.4$ and {${\tau_{frac}=0.3}$}.]{
	\includegraphics[]{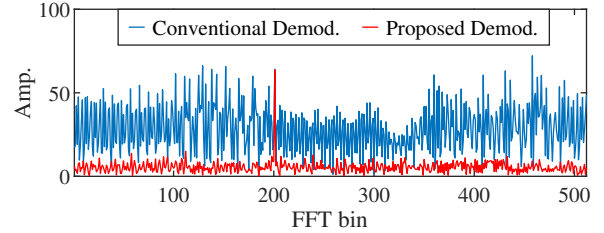}
    \label{fig:FFT2}}
    
    \subfloat[The interference signal includes  $\chi$ and integer $\tau$, where $\chi_{frac}=0.45$.]{
	\includegraphics[]{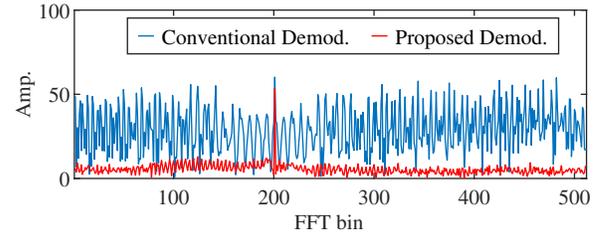}
    \label{fig:FFT3}}
        
    \subfloat[The interference signal includes $\tau$ and integer $\chi$, where $\tau_{frac}=-0.45$.]{
	\includegraphics[]{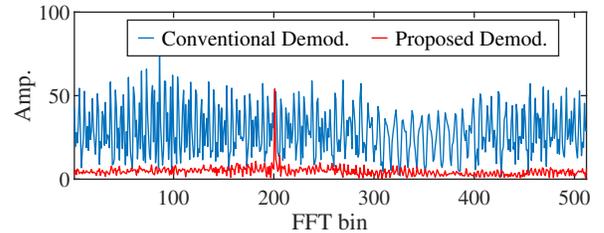}
    \label{fig:FFT4}}
    
\caption{Comparison of different demodulation methods in terms of DFT spectrum under noise-free condition. A synchronized LoRa signal with $\mathrm{SF}_u$=9 interfered with non-aligned interference with $\mathrm{SF}_i$=7 under SIR=-20dB is demodulated using conventional demodulation method (blue curve). The same signal is demodulated after interference cancellation with $d$=8 (red curve). }
    \label{FFT_spectrum}
\end{figure}

After reconstructing the interference using $\widehat l_m$ and $\widehat A^m_i$, the interference cancellation is performed using subtraction. Finally, the conventional demodulation method can be used for recovering the transmitted symbol from the desired user. A comparison between the proposed method and the conventional demodulation in terms of the DFT spectrum of the desired signal is presented. First, Figure \ref{fig:FFT1} shows the DFT spectrum of a LoRa signal with inter-SF interference, where the interference includes only integer CFO and STO. It can be recognized that without interference cancellation the peak from the desired symbol cannot be observed due to the interference, whereas after interference cancellation the desired peak becomes prominent. {Furthermore, Figure \ref{fig:FFT2}, \ref{fig:FFT3} and \ref{fig:FFT4} illustrate the DFT spectrum for the same signal, but also with fractional CFO and STO in the interference. It can be noticed that the interference can still be accurately suppressed in spite of the fractional frequency component and the sub-sample offset.} 
	
\section{Simulation Results}\label{Sec:Simulation}
In order to evaluate the validity of the proposed approach, we resort to numerical simulations. The LoRa performance is evaluated under two different channel conditions, namely noiseless and AWGN. A LoRa transceiver chain is simulated following the system model presented in Section \ref{sec:System_model}. For each simulation run, we create an overlapped signal by adding the desired signal modulated with $\mathrm{SF}_u$ to a number of random interfering symbols with $\mathrm{SF}_i$. Firstly, the performance of $\widehat{\mathrm{SF}_i}$ identification is evaluated. Then, the selection of the number of segments $d$ is also investigated to obtain the best performance of interference estimation. Afterwards, the LoRa performance using the proposed scheme is shown. Similarly to \cite{croce2017impact}, we assume that the desired signal is perfectly synchronized, while the interference is randomly shifted in time. The LoRa demodulation performance in terms of both SER and BER is presented. In the end, the proposed demodulation under interference-free condition is also investigated. Simulation results show that the influence caused by interference can be neglected under SIR $>-10$ dB, which is in accordance with \cite{croce2017impact} and \cite{Interference_analysis}. Therefore, we only consider SIR$\leqslant-10$dB. Unless specified otherwise, the settings for the simulations are listed in Table \ref{tab:sim}.
\begin{table}[t!]
\centering
\captionsetup{justification=centering}
\caption{Simulation parameters.
}
\begin{tabular}{@{}ll@{}}
\toprule
Parameter          &Value \\ \midrule
$\mathrm{SF}_u$ &9, 8     \\
$\mathrm{SF}_i$&7, 10      \\
Coding rate   &4/7     \\
Integer $\tau$ &$\lbrack0,N_i-1\rbrack$\\
Fractional $\tau$ &$\lbrack-0.5,0.5\rbrack$\\
Integer $\chi$ &$\lbrack-N_i/4,N_i/4-1\rbrack$\\
Fractional $\chi$ &$\lbrack-0.5,0.5\rbrack$\\\bottomrule
\end{tabular}\label{tab:sim}
\end{table}
\subsection{Performance of interference SF identification}
It can be concluded from Section \ref{sec:iden} that the accuracy of interference SF identification depends on the noise and interference level. Since the proposed SF identification scheme is based on the demodulation of the interference, a low SIR increases the accuracy of estimation. Therefore, the interference SF estimation is performed under harsh conditions with high SIR and low SNR. Table \ref{tab:iden} shows the confusion matrices of interference SF identification with $\mathrm{SF}_u=9$ under SIR=-10dB and SNR=-20dB. The transmitter is assumed 
to be synchronized with the receiver, while
the interference signals are randomly shifted in time for non-alignment. Moreover, for each simulation run the interference signal includes random $\tau$ and $\chi$ with the range in Table \ref{tab:sim}. Each confusion matrix includes the results of all possible interference SFs with a fixed $N_s$. It can be easily recognized that the larger the $N_s$, the higher the identification accuracy since more signal data are processed. Furthermore, a smaller interference SF is more difficult to identify, since it is more sensitive to noise due to the associated smaller processing gain. As shown in Table \ref{tab:512}, the identification accuracy of $\mathrm{SF}_i \geq 11$ is 1 while the accuracy is only 0.684 of $\mathrm{SF}_i=7$. As $N_s$ goes from 512 to 8192, the identification accuracy of $\mathrm{SF}_i=7$ increases to 0.962. As $N_s$ further increases, the accuracy will be closer to 1.
\begin{table*}[ht]
\captionsetup{justification=centering}
\centering
\caption{Confusion matrix of interference SF identification accuracy with $\mathrm{SF}_u=9$ under SIR=-10dB and SNR=-20dB.}
\setlength{\tabcolsep}{3pt}
\subfloat[$N_s$=512.]{\begin{tabular}{|c|*{5}{c|}}
\hline 
\diagbox{$\mathrm{SF}_i$}{$\widehat {\mathrm{SF}}_i$}   & 7 & 8 & 10 & 11 & 12 \\ \hline
7  &\col0.684   &0.039   &0.071    &0.098    &0.108    \\ \hline
8  &0.011   &\col0.904   &0.022   &0.030    &0.033    \\ \hline
10 &0   &0   &\col0.998    &0.001    &0.001    \\ \hline
11 &0   &0   &0    &\col1   &0    \\ \hline
12 &0   &0   &0    &0    &\col1    \\ \hline
\end{tabular}\label{tab:512}}
\quad
\subfloat[$N_s$=2048.]{\begin{tabular}{|c|*{5}{c|}}
\hline 
\diagbox{$\mathrm{SF}_i$}{$\widehat {\mathrm{SF}}_i$}   & 7 & 8 & 10 & 11 & 12 \\ \hline
7  &\col0.880   &0.026   &0.024    &0.024    &0.046    \\ \hline
8  &0.003   &\col0.988   &0.002   &0.002    &0.005    \\ \hline
10 &0   &0   &\col1    &0    &0    \\ \hline
11 &0   &0   &0    &\col1   &0    \\ \hline
12 &0   &0   &0    &0    &\col1    \\ \hline
\end{tabular}}
\quad
\subfloat[$N_s$=8192.]{\begin{tabular}{|c|*{5}{c|}}
\hline 
\diagbox{$\mathrm{SF}_i$}{$\widehat {\mathrm{SF}}_i$}   & 7 & 8 & 10 & 11 & 12 \\ \hline
7  &\col0.962   &0.01   &0.01    &0.01    &0.008    \\ \hline
8  &0.001   &\col0.999   &0   &0    &0    \\ \hline
10 &0   &0   &\col1    &0    &0    \\ \hline
11 &0   &0   &0    &\col1   &0    \\ \hline
12 &0   &0   &0    &0    &\col1    \\ \hline
\end{tabular}}
\label{tab:iden}
\end{table*}
\subsection{Investigation on the required number of segments}
\begin{figure}[t!]
    \centering
    \subfloat[$\mathrm{SF}_u$=9, $\mathrm{SF}_i$=7.]{
        \includegraphics[width=0.24\textwidth]{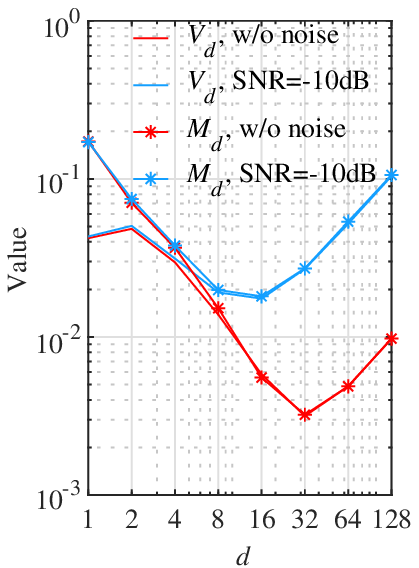}
    }
    \subfloat[$\mathrm{SF}_u$=8, $\mathrm{SF}_i$=10.]{
	\includegraphics[width=0.24\textwidth]{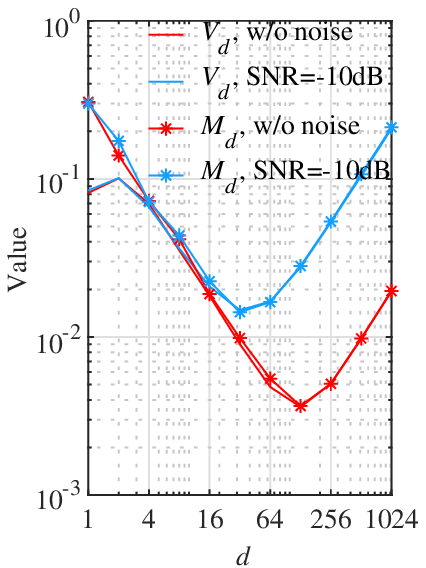}
    }
    \caption{Influence of $d$ on the variance and MSE of the estimated amplitude of interference signal for different combinations of $\mathrm{SF}_u$ and $\mathrm{SF}_i$ with SIR=-20dB under noise-free condition and AWGN, respectively.}
    \label{fig:d_influence}
\end{figure}

From Section \ref{sec:symbol_value}, we know that as the number of segments increases, the accuracy of the estimation of the interference signal improves. However, the DFT magnitude is inversely proportional to $d$, and a low peak will be more likely disturbed by noise, thus degrading the estimation performance. For investigating the influence of $d$ on the estimation of the interference signal, {the mean square error (MSE) $M_d$ of $|\widehat A_i^m|_d$ is calculated, where $|\widehat A_i^m|_d$ is the magnitude of $\widehat A^m_i$ using $d$ segments. The minimum of $M_d$ can only approximately achieve the best estimation performance \cite{kay1993fundamentals}, since the proposed estimation is valid for same-symbol segments but inter-symbol segments are present. Nevertheless, this inter-symbol segments have negligible influence on the interference signal reconstruction. Figure \ref{fig:d_influence} presents $M_d$ versus $d$ under both noise-free and AWGN, where fractional $\tau$ and $\chi$ are not included. It can be seen that for both cases $M_d$ first reaches its lowest point as $d$ increases to a certain value $d_{best}$, which represents the best performance of interference signal estimation, and then rebounds. Since segments with shorter length result in lower processing gain, we observe the performance degradation of interference estimation. It can be also recognized that, $d_{best}$ under AWGN is smaller than the one obtained under noise-free condition, which verifies the previous consideration that smaller segments are more sensitive to noise. Analysis on all SF combinations also shows that the $d$ values minimizing $M_d$ are identical for the same $\mathrm{SF}_i$. In other words, the SF of the desired signal has little impact on the estimation of the interference. Although $M_d$ can indicate $d_{best}$, it cannot be calculated in real wireless condition. Therefore, $V_d=\mathrm{Var}\left ( |\widehat A^m_i|_d\right)$ is also illustrated in Figure \ref{fig:d_influence}, which is the variance of $ |\widehat A^m_i|_d$. It is clear that $V_d$ and $M_d$ are very close and coincide after $d=4$, which indicates a small bias of the proposed estimation for $d<4$. Moreover, from both curves the same $d_{best}$ is obtained. Thus, the required number of segments can be determined from $V_d$ before interference cancellation. First, we loop through all possible $d$ based on $\mathrm{SF}_i$ to calculate $V_d$ for each $d$. Then, $d_{best}$ is selected from the smallest $V_d$ which can be expressed as
\begin{equation}\label{eq:dbest}
    d_{best}=\arg\min_{d\in\mathbb{D}}V_d.
\end{equation}}
\subsection{Demodulation performance}
\begin{figure}[ht]
\vspace{-.5cm}
    \centering
     \subfloat[$\mathrm{SF}_u$=9, $\mathrm{SF}_i$=7.]{
        \includegraphics[]{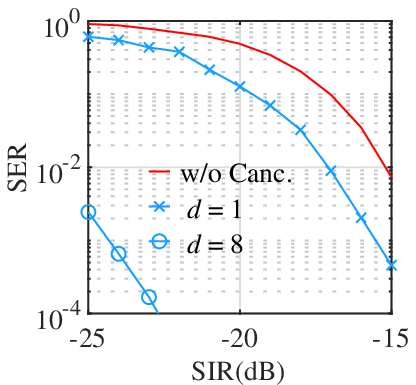}
    }
    \subfloat[$\mathrm{SF}_u$=8, $\mathrm{SF}_i$=10.]{
	\includegraphics[]{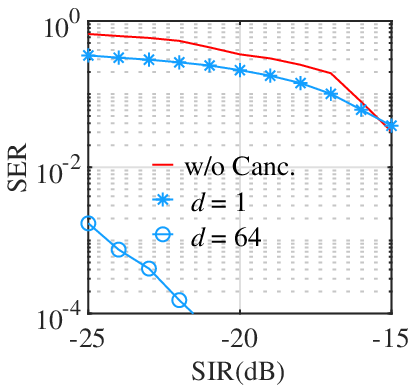}
    }
    \caption{SER results of the proposed demodulation and the conventional demodulation for different combinations of $\mathrm{SF}_u$ and $\mathrm{SF}_i$ in function of SIR under noise-free condition. Different $d$ are chosen for the proposed demodulation. The desired signal is perfectly synchronized.}
    \label{fig:simu_diff_d}
\end{figure}

Figure \ref{fig:simu_diff_d} shows the SER versus SIR under noise-free condition for the perfectly synchronized desired signal, with a smaller and a larger $\mathrm{SF}_i$ compared to $\mathrm{SF}_u$, respectively. $N_s$ is set to 8192 for $\mathrm{SF}_i$ estimation. It can be recognized that the proposed method significantly reduces the SER compared with the conventional demodulation. Above all, we can see that the results with $d=1$, which is interference cancellation without segmentation, already improve SER performance. Then, the SER further decreases as $d$ increases.  When $d$ is close to $d_{best}$, the SER is already close to zero.
\begin{figure}[ht]
    \centering
   \includegraphics[]{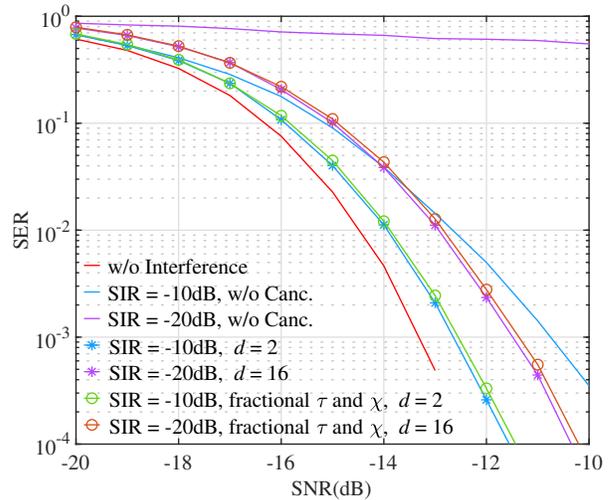}
   \caption{SER results of the proposed demodulation and the conventional demodulation in function of SNR under AWGN with $\mathrm{SF}_u$=9 and $\mathrm{SF}_i$=7. The desired signal is perfectly synchronized.}
   \label{fig:Sim2_1}
\end{figure}
 Figure \ref{fig:Sim2_1} shows the SER of the synchronized frame under AWGN with $\mathrm{SF}_u$=9, $\mathrm{SF}_i$=7, where the curves of the proposed method are obtained with $d_{best}$. $N_s$ is set to 8192 for $\mathrm{SF}_i$ identification. The red solid line describes the SER result under interference-free condition using conventional demodulation. We can also recognize from other unmarked curves that the interference degrades the SER performance of the conventional demodulation significantly. When SIR$=-20$dB, the SER is above 0.5 for each SNR. It can be observed from the marked curves that the proposed method still works in AWGN and outperforms the conventional method. Furthermore, the demodulation performance of the proposed method only shows negligible degradation when the interference includes random fractional CFO and STO with the range in Table \ref{tab:sim}. Most importantly, it is able to avoid decoding failure under low SIR condition. In addition, we notice that $d_{best}$ becomes smaller as the SIR increases, since a large $d$ will degrade the interference estimation performance when the noise is more dominant than the interference, leading to inaccurate interference cancellation. 
 
\begin{figure}[t!]
     \centering
     \includegraphics{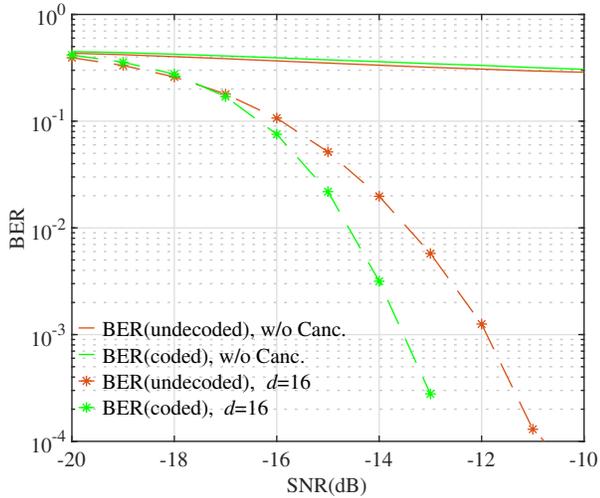}
     \caption{Undecoded and coded BER results in function of SNR under AWGN with $\mathrm{SF}_u$=9, $\mathrm{SF}_i$=7 and SIR=-20dB. Both proposed demodulation and conventional demodulation are performed. The desired signal is perfectly synchronized.}
     \label{fig:BER}
 \end{figure}

 Furthermore, Figure \ref{fig:BER} shows the BER of the synchronized frame under AWGN with $\mathrm{SF}_u$=9, $\mathrm{SF}_i$=7 and SIR=-20 dB. The LoRa PHY coding is performed following the steps in \cite{AfCoded}. First, the payload bits are encoded by (7, 4) Hamming code that can correct single-bit errors. Second, the codewords are interleaved using diagonal interleaver for more robustness to burst errors. Then, Gray indexing is employed for mitigating bit errors induced by off-by-one data symbol errors. In order to show the impact of coding on the BER, both coded and undecoded BER are presented. It can be first noticed from the two solid lines that the coding cannot reduce the BER  using conventional LoRa demodulation under low SIR. While after interference cancellation, both undecoded and coded BER are reduced. Notably, combining interference cancellation and LoRa PHY coding can further improve the demodulation performance.
 
\subsection{Interference-free condition}
\begin{figure}[t!]
    \centering
    \includegraphics{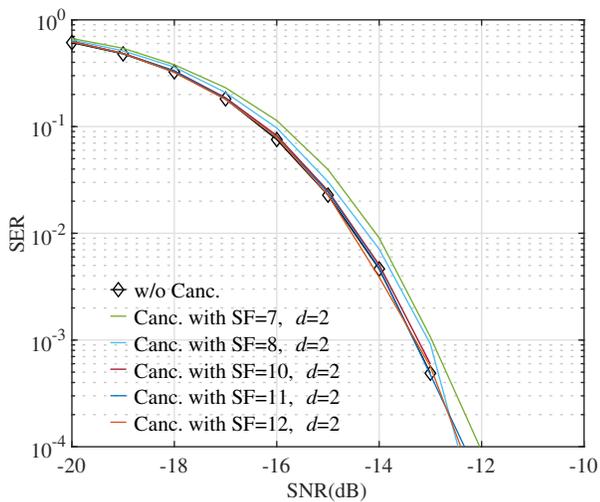}
    \caption{SER of interference-free scenario with $\mathrm{SF}_u$=9 under AWGN, where the interference cancellation is wrongly performed. The desired signal is perfectly synchronized.}
    \label{fig:nointer}
\end{figure}

It can be seen from Algorithm \ref{al:id} that the scheme will always generate an identified $\mathrm{SF}_i$ whether interference exists or not. When no interference exists, the proposed scheme will cause false alarms and still perform interference cancellation using the wrongly identified $\mathrm{SF}_i$. Figure \ref{fig:nointer} illustrates the SER of the interference-free scenario using $\mathrm{SF}_u$=9 under AWGN. The interference cancellation is performed for all possible $\mathrm{SF}_i$ with $d=2$. It can be seen that the false interference cancellation slightly degrades the LoRa demodulation performance and the smaller the wrongly identified $\mathrm{SF}_i$, the higher the SER. Moreover, this impact can be ignored at high SNR since the SER of all the wrongly identified $\mathrm{SF}_i$ is closer to zero under SNR $>$ -12dB. Therefore, even under interference-free condition, where the interference cancellation is wrongly performed, the proposed scheme still can be applied at the gateway, with a little loss of demodulation accuracy.

\section{Conclusion}\label{Sec:conclu}
In this work, the non-aligned inter-SF interference from one interferer is modeled and analyzed. A receiver is structured to reconstruct the interference signal and cancel it from the received signal. The demodulation of the desired signal after interference cancellation is performed under noiseless condition and AWGN, respectively. Simulation results show that the demodulation performance is significantly improved compared to the commonly used method for both conditions. Furthermore, the proposed scheme is also robust against STO and CFO in the interference signal. With the proposed method, the near-far problem caused by inter-SF interference is mitigated, which can offer more options for SF allocation in LoRa networks. Consequently, the scalability of LoRa can be enhanced.

{As LoRa synchronization relies on DFT-based demodulation of the preambles, the detection and synchronization of the desired signal can be also influenced by the inter-SF interference. The probability of detection can be efficiently improved by using raw up-chirps as many as possible to equivalently increase the SIR. While the synchronization symbols and the down-chirps are more sensitive to low SIR due to their small number, degrading the synchronization performance. Performing interference cancellation before synchronization can be a choice to solve this problem.}

Indeed, this scheme is also applicable to the case in which interference does not exist, at the expense of more computation consumption at the gateway. Future works should explore the false alarms caused by the algorithm of interference SF identification, where a threshold is required to indicate the existence of the interferer, and even the coexistence of multiple interferers to perform iterative interference cancellation.

\bibliographystyle{ieeetr}
\bibliography{references.bib}

\vskip -2\baselineskip plus -1fil
\begin{IEEEbiography}[{\includegraphics[width=1in,height=1.25in,clip,keepaspectratio]{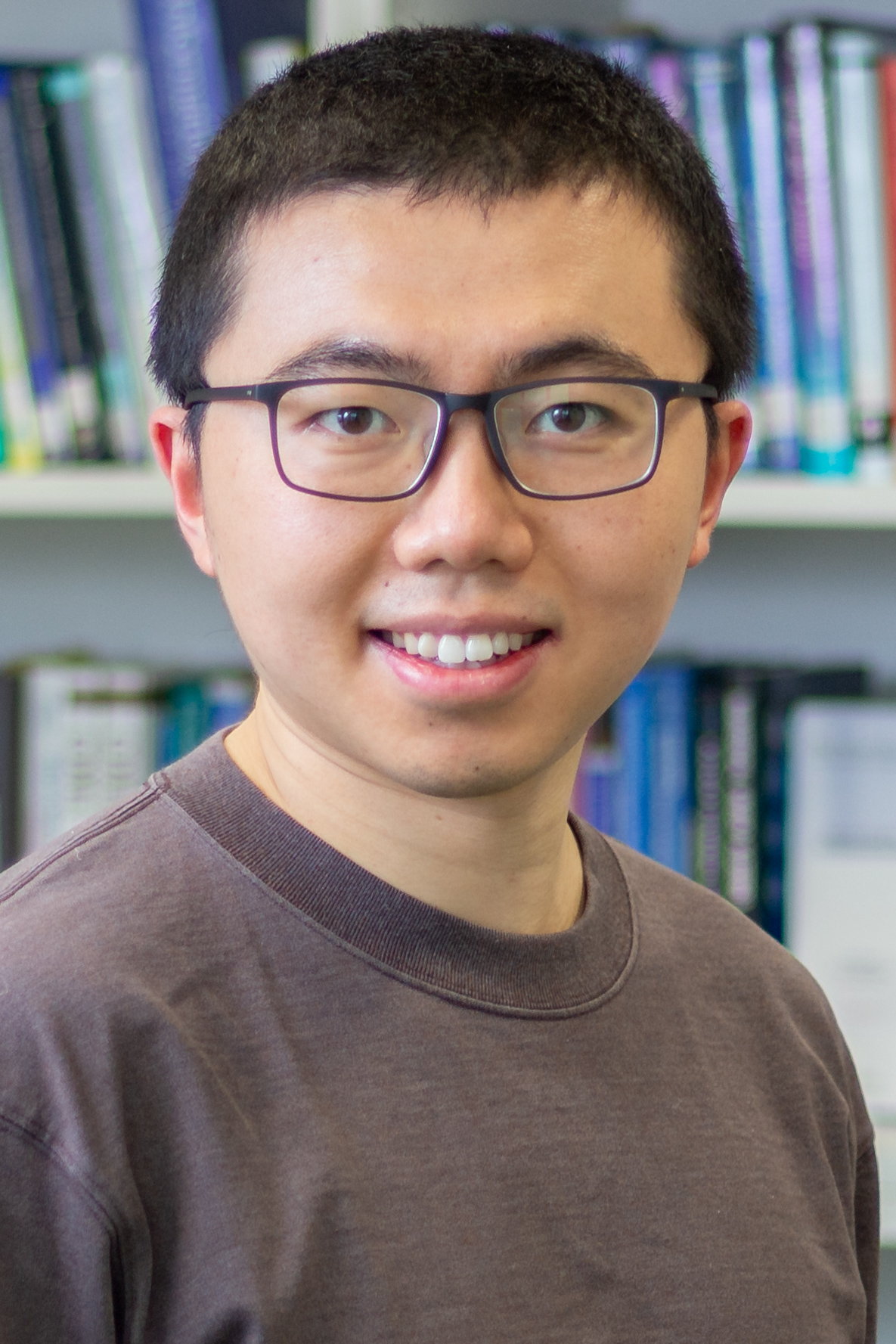}}]{Qiaohan Zhang}received  the
Diploma (Dipl. Ing.) degree in electrical engineering in 2021 from the Fakult\"at Elektrotechnik und Informationstechnik, Technische Universit\"at Dresden (TUD), Germany. Currently, he is a research associate in Vodafone Chair Mobile Communications Systems, Technische Universität Dresden, Germany from January 2022. His research interests include signal detection and synchronization, machine learning, and chirp spread spectrum communications.
\end{IEEEbiography}

\vskip -2.5\baselineskip plus -1fil	
\begin{IEEEbiography}[{\includegraphics[width=1in,height=1.25in,clip,keepaspectratio]{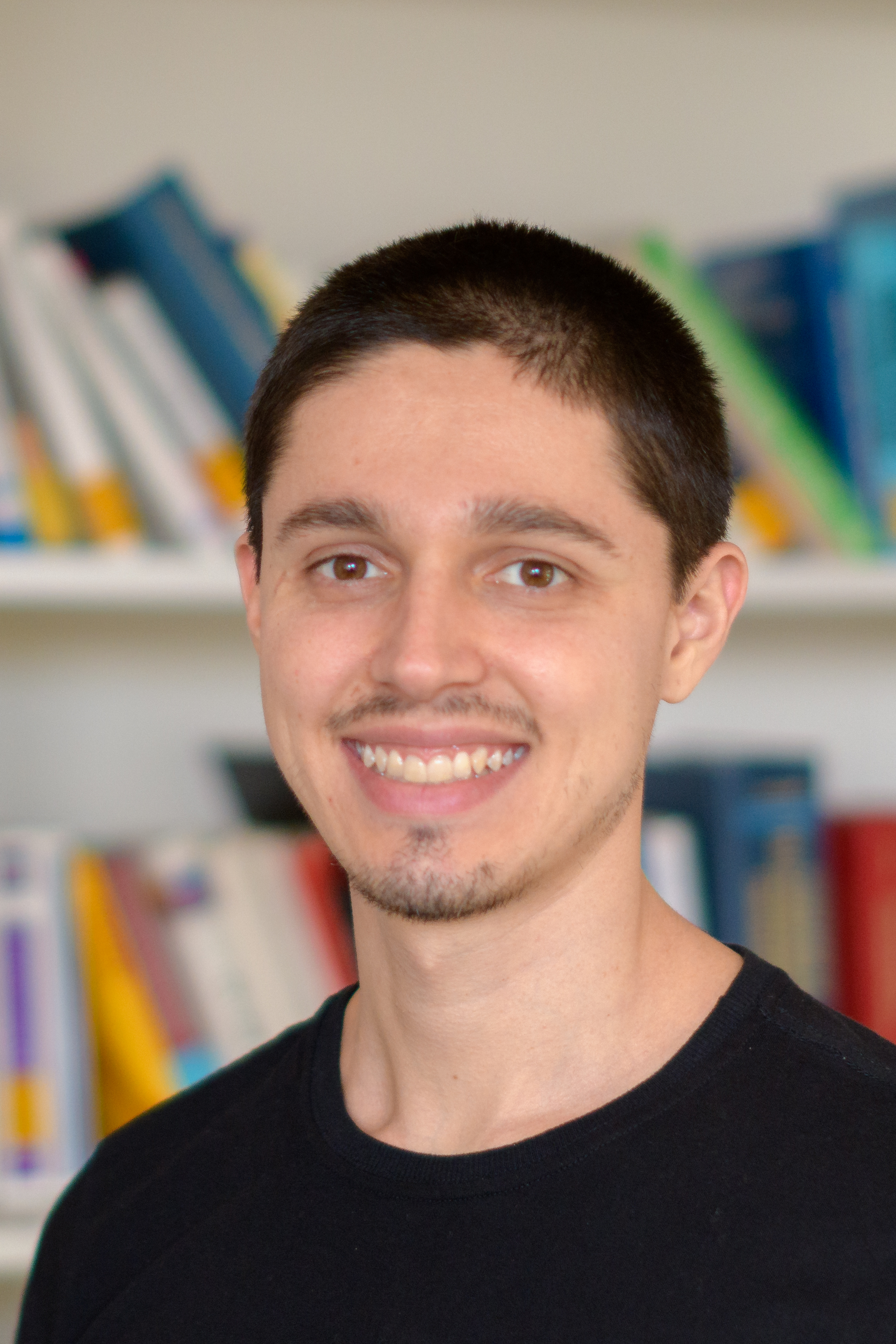}}]{Ivo Bizon}
	received the five-year BSc. and MSc. degrees in Electrical Engineering from the Instituto Nacional de Telecomunica\c{c}\~oes (Inatel), Brazil in 2016 and 2018, respectively. 
	Currently he is pursuing his PhD at Technische Universi\"at Dresden (TUD), Germany, and working as a research associate at the Vodafone Chair Mobile Communications Systems. His current research interests include deep learning based localization schemes and modulation techniques for future long range and low power wireless systems.
\end{IEEEbiography}

\vskip -2\baselineskip plus -1fil
\begin{IEEEbiography}[{\includegraphics[width=1in,height=1.25in,clip,keepaspectratio]{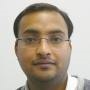}}]{Atul Kumar }received the B.Tech. degree in electronics and communication engineering, in 2013, and the M.S. degree in electronics engineering in September 2015, and the Ph.D. degree in information engineering at the Dipartimento di Elettronica, Informazione, and Bioingegneria in December 2018 from the Politecnico di Milano, Milan, Italy. From 2018 to 2021, he was with the Vodafone Chair at TU Dresden, where he was a research associate. He is currently an Assistant Professor in the Department of Electronics Engineering, IIT(BHU) Varanasi, India. Since 2017, he was the Director of AtlaMedico TechSolutions Pvt Ltd, he founded the company for the development of a wireless medical device for intensive care units, India, a technology start-up for the design, optimization, and operation of the medical device. His main research interests include 6G, wireless cellular systems, synchronization errors, joint sensing and communication technology, AI techniques for end-to-end (E2E) prediction of critical Quality-of-Service (QoS), New Radio, massive MIMO (mMIMO), orthogonal frequency division multiplexing, generalized frequency division multiplexing, Internet of things, and Narrowband power line commutations.
\end{IEEEbiography}

\vskip -2\baselineskip plus -1fil
\begin{IEEEbiography}[{\includegraphics[width=1in,height=1.25in,clip,keepaspectratio]{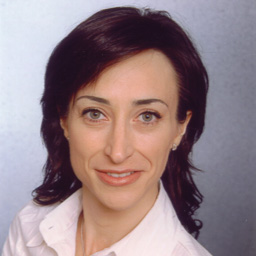}}]{Ana Belen Martinez}
received the M.Sc. degree in electrical engineering in 2012 from the Fakult\"at Elektrotechnik und Informationstechnik, Technische Universit\"at Dresden (TUD), Germany. Currently she is pursuing her PhD at the Vodafone Chair Mobile Communications Systems, TUD. Her main research interests include digital signal processing for wireless communications systems with focus on estimation, detection and synchronization. She has several years of teaching experience and works actively in the area of mobile campus networks.
\end{IEEEbiography}

\vskip -2\baselineskip plus -1fil
\begin{IEEEbiography}[{\includegraphics[width=1in,height=1.25in,clip,keepaspectratio]{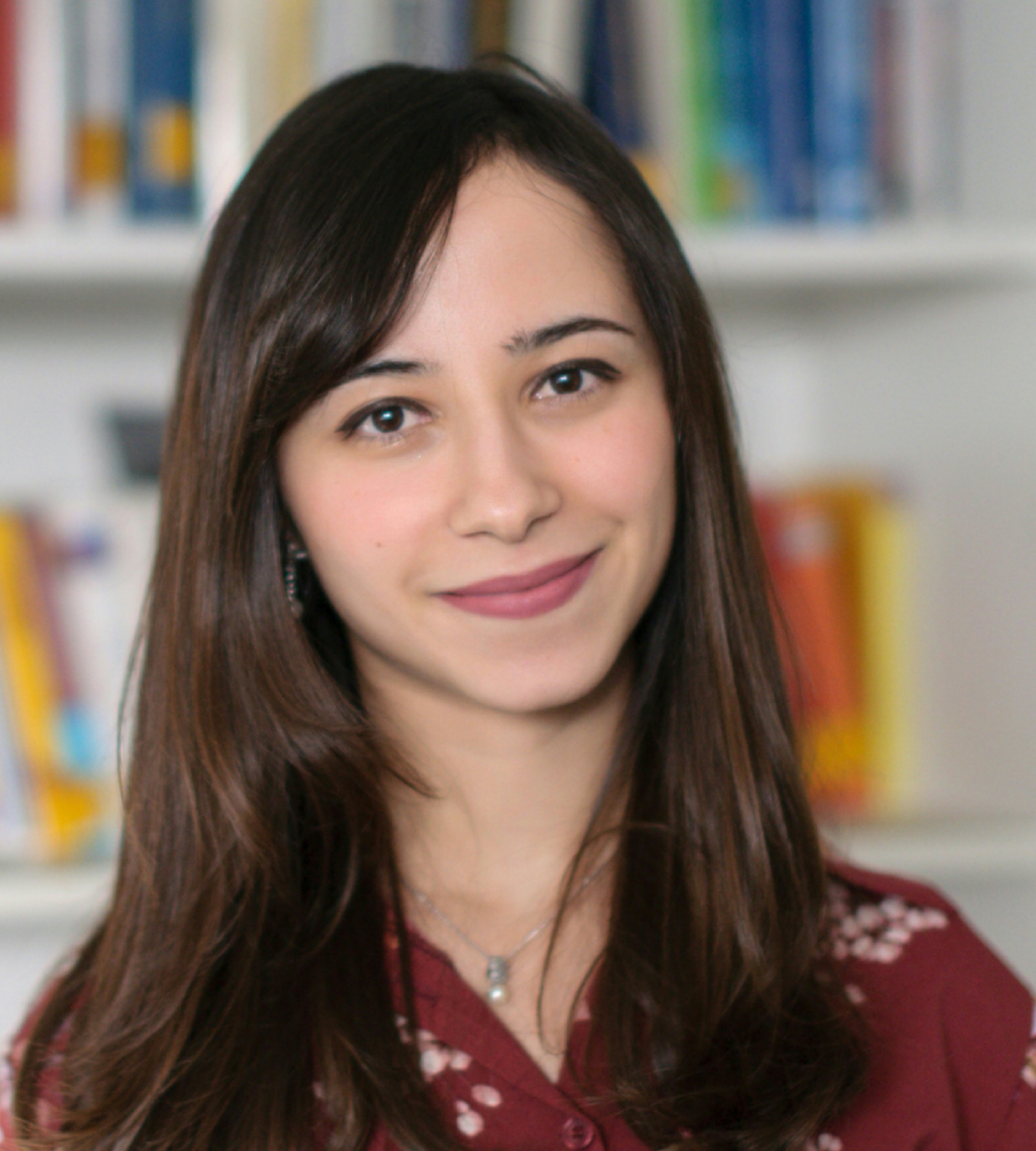}}]{Marwa Chafii}
	Marwa Chafii received her Ph.D. degree in electrical engineering in 2016, and her Master's degree in the field of advanced wireless communication systems (SAR) in 2013, both from CentraleSupélec, France. Between 2014 and 2016, she has been a visiting researcher at Poznan University of Technology (Poland), University of York (UK), Yokohama National University (Japan), and University of Oxford (UK). She joined the Vodafone Chair Mobile Communication Systems at the Technical University of Dresden, Germany, in February 2018 as a research group leader. Since September 2018, she is and an associate professor at ENSEA, France, where she holds a Chair of Excellence on Artificial Intelligence from CY Initiative. She received the prize of the best Ph.D. in France in the fields of Signal, Image \& Vision, and she has been nominated in the top 10 Rising Stars in Computer Networking and Communications by N2Women in 2020. Since 2019, she serves as Associate Editor at IEEE Communications Letters where she received the Best Editor Award in 2020. She is currently vice-chair of the IEEE ComSoc ETI on Machine Learning for Communications, leading the Education working group of the ETI on Integrated Sensing and Communications, and research lead at Women in AI.
\end{IEEEbiography}

\vskip -2\baselineskip plus -1fil
\begin{IEEEbiography}[{\includegraphics[width=1in,height=1.25in,clip,keepaspectratio]{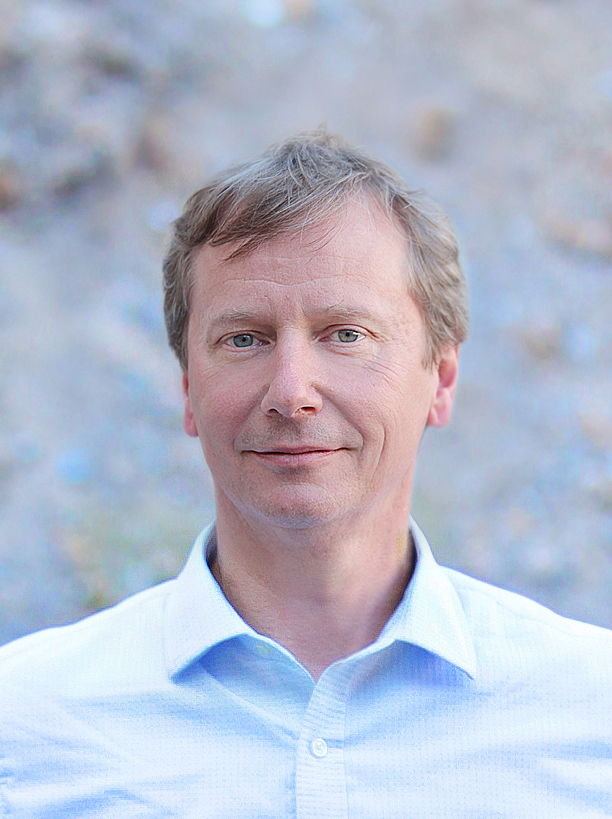}}]{Gerhard Fettweis}
	is Vodafone Chair Professor at TU Dresden since 1994, and heads the Barkhausen Institute since 2018, respectively. He earned his Ph.D. under H. Meyr's supervision from RWTH Aachen in 1990. After one year at IBM Research in San Jose, CA, he moved to TCSI Inc., Berkeley, CA. He coordinates the 5G Lab Germany, and 2 German Science Foundation (DFG) centers at TU Dresden, namely cfaed and HAEC. His research focusses on wireless transmission and chip design for wireless/IoT platforms, with 20 companies from Asia/Europe/US sponsoring his research.
	Gerhard is IEEE Fellow, member of the German Academy of Sciences (Leopoldina), the German Academy of Engineering (acatech), and received multiple IEEE recognitions as well has the VDE ring of honor. In Dresden his team has spun-out sixteen start-ups, and setup funded projects in volume of close to EUR 1/2 billion. He co-chairs the IEEE 5G Initiative, and has helped organizing IEEE conferences, most notably as TPC Chair of ICC 2009 and of TTM 2012, and as General Chair of VTC Spring 2013 and DATE 2014. 
\end{IEEEbiography}

\end{document}